# A Sub-block Based Image Retrieval Using Modified Integrated Region Matching

E R Vimina[1] and K Poulose Jacob[2]

[1] **Department of Computer Science, Rajagiri College of Social Sciences**
**Kochi, Kerala, India**
*Vimina_er@yahoo.com*

[2] **Department of Computer Science, Cochin University of Science and Technology**
**Kochi, Kerala, India**
*kpj@cusat.ac.in*

**Abstract**

This paper proposes a content based image retrieval (CBIR) system using the local colour and texture features of selected image sub-blocks and global colour and shape features of the image. The image sub-blocks are roughly identified by segmenting the image into partitions of different configuration, finding the edge density in each partition using edge thresholding, morphological dilation. The colour and texture features of the identified regions are computed from the histograms of the quantized HSV colour space and Gray Level Co- occurrence Matrix (GLCM) respectively. A combined colour and texture feature vector is computed for each region. The shape features are computed from the Edge Histogram Descriptor (EHD). A modified Integrated Region Matching (IRM) algorithm is used for finding the minimum distance between the sub-blocks of the query and target image. Experimental results show that the proposed method provides better retrieving result than retrieval using some of the existing methods.

## 1. Introduction

Content Based Image Retrieval (CBIR) has become an important area of research with the ever increasing demand and use of digital images in various fields such as medicine, engineering, sciences, digital photography etc. Unlike the traditional method of text-based image retrieval in which the image search is based on textual description associated with the images, CBIR systems retrieve images based on the content of the image such as colour, texture, shape or any other information that can be automatically extracted from the image itself and using it as a criterion to retrieve content related images from the database. The retrieved images are then ranked according to the relevance between the query image and images in the database in proportion to a similarity measure calculated from the features [1][2][3].

## 2. Related Work

Of the many variants of CBIR systems, query-by-example (QBE) is the most widely supported method. Here the user formulates the query by giving an example image. The features of this query image will be extracted and compared with the pre-extracted features of the images in the database and the most similar images will be returned to the user. Most of the early CBIR systems rely on global features of the query image to retrieve similar images [4][5][6][15]. But they more often fail either due to the lack of higher-level knowledge about what exactly was of interest to the user in the query image or due to the fact that global features cannot sufficiently capture the important properties of individual objects. Recently, much research has focused on region-based techniques [2][3][7][16][19][31]. Such systems can be classified into two types, the ROI defined by the user or ROI identified by machine learning methods. In the first type the user can randomly select the region of the image based on his or her need and search for similar images [16][31]. Although this method captures meaningful object regions, sometimes it is a tedious and boring task for the user. The second type either subdivide the image into fixed blocks [19][20][21] or partition the image into different meaningful regions using segmentation algorithms [2][3][7][23]. Performance of segmentation based methods depends highly on the quality of the segmentation as the average features of all pixels in a segment are often used as the features of that segment. Small areas of incorrect segmentation might make the representation very different from that of the real object. Incorrect segmentation may also affect the shape features. Also accurate segmentation is still a challenging problem and the computational load of segmentation method is heavier. For the fixed block segmentation methods the computational cost is less and also provides satisfactory results comparable with that of the pixel-wise segmentation methods even if the objects are not segmented correctly. Some other CBIR systems [16] [30] extract salient points (also known as interest points) [28] [29], which are locations in an image where there is a significant variation with respect to a chosen image feature. In salient point based methods, feature vector is created for each salient point and the selection of the number of salient points is very important. These representations enable a retrieval method to have a representation of different local regions of the image, and thus these images can be searched based on their local characteristics.

## 3. Proposed Method

In the proposed method fixed block segmentation is used. The images are divided into different sized blocks for feature extraction. Feature vectors are extracted from selected grids of different configurations (3x3 grid, horizontal and vertical grids, central block and the entire image) (Fig.1). Unlike some block based image retrieval systems that uses all the sub-blocks for feature extraction and similarity measurement, our system uses selected blocks only reducing the computational time and cost.

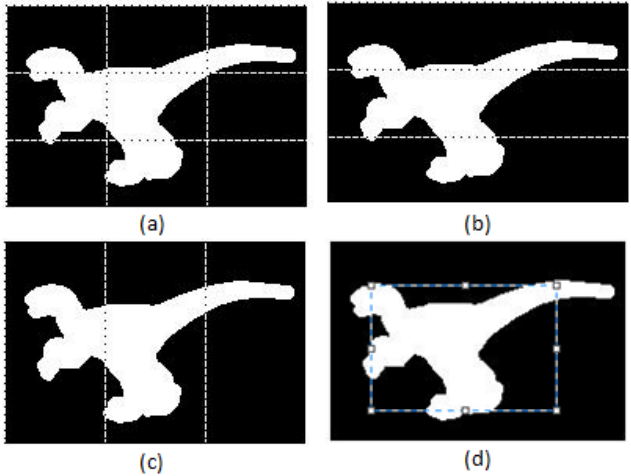

Fig. 1. Different image configurations for feature extraction

### 3.1 Attention Center and Central Block Extraction

To find the attention center of an image, the first step is to find the salient regions. In an image all regions may not be important or perceptually salient. When an image is mapped into the appropriate feature space salient regions will stand out from the rest of the data and can more easily be identified. To identify the salient regions the images are initially cropped by 20 pixels in the horizontal and vertical direction from the border in-order to avoid the effect of unwanted edges in the border regions. The resultant image is then converted to gray scale and blurred with Gaussian filter to discard noise. The canny edge filter is used for extracting the prominent edges. Center of mass (centroid) of the resultant image is found and is termed as attention center.

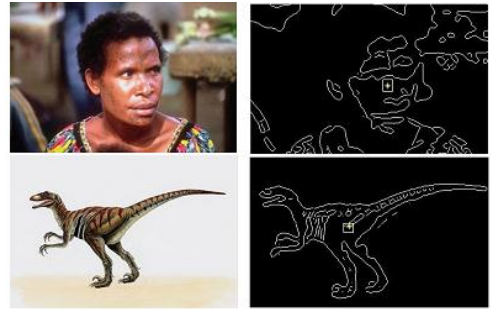

Fig. 2. Original image (Left) and the edge image marked with attention center (Right)

The rectangular region around the attention center with dimensions half the size of the original image is taken as the center block.

### 3.2 Sub-block Selection

To identify the sub-blocks /object regions, first the grayscale image is computed and edge map is detected using Sobel edge filter with a threshold value of τ (τ <1 so that the edges are boosted). The gaps in the edge map are bridged by dilating it with 'line' structuring element, that consists of three 'on' pixels in a row, in the 0, 45, 90 and 135 directions. The holes in the resultant image are then filled to get the approximate location of the objects. The objects are identified correctly if the background is uniform.

A sub-block is selected for further processing, feature extraction and is identified as region of interest (ROI) if τ'% of the sub-block is part of the object region. Ie, if the number of white pixels in that sub-block is τ'% of the sub-block with maximum white pixel density, it is identified as a region of interest. For example, for the 3x3 partitioned image in Fig.3, regions 1, 3, 4, 5, and 8 are the ROIs. Only these sub-blocks take part in further computations for calculating the similarity along with the global colour and shape features of the entire image [26]. The horizontal and vertical ROIs are also identified in the same manner

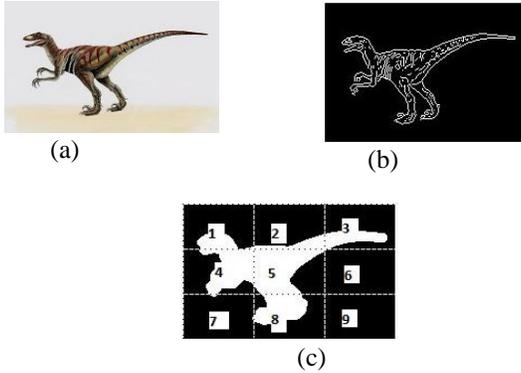

Fig. 3. (a)Original image (b)Edge map after sobel edge filtering (c) Edge map after edge thresholding and morphological dilation

## 4. Feature Extraction

The colour and texture features of the selected sub-blocks are extracted for similarity computation between the query and the candidate images in the database. Global colour and shape features are also computed for this purpose.

### 4.1 Colour

Colour features are extracted using the histograms of HSV colour space. For this purpose, the HSV colour space is quantized into 18 bins of Hue, 3 bins of Saturation and 3 bins of Value. The histogram of each of these channels are extracted resulting in a 24 dimensional colour feature vector that is normalized in the range of [0,1]. For each image both global and local colour features are extracted.

### 4.2 Texture

Texture features are extracted using the Gray Level Co-occurrence Matrix (GLCM). It is a matrix showing how often a pixel with the intensity (gray-level) value i occurs in a specific spatial relationship to a pixel with the value j. It is defined by $P(i,j| d,\theta)$, which expresses the probability of the couple of pixels at $\theta$ direction and d interval. Once the GLCM is created various features can be computed from it. The most commonly used features are contrast, energy, entropy, correlation and homogeneity. We have taken $d=1$ and $\theta = 0^o, 45^o, 90^o$ and $135^o$ for computing the texture features. Contrast, energy, correlation and homogeneity are taken in all the four directions and entropy of the whole block is separately calculated as it gave better retrieving results. Thus 17 texture feature vectors are calculated for each sub-block.

### 4.3 Shape

Shape feature provide important semantic information due to human's ability to recognize objects through their shape. However, this information can only be extracted by means of a segmentation similar to the one that the human visual system implements which is still a challenging problem. Here Edge Histogram Descriptor (EHD) is used for shape feature extraction[13][14]. It represents the local edge distribution of the image by dividing image space into 4× 4 sub-images and representing the local distribution of each sub-image by a histogram. For this, edges in the sub-images are categorized into five types; vertical, horizontal, 45-degree diagonal, 135-degree diagonal and non-directional edges (Fig.4). The edge histogram for the sub-images are computed resulting in a shape feature vector of size 80.

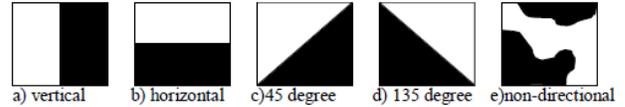

Fig 4 Five types of edges in the Edge Histogram Descriptor

## 5. Similarity Computation

The $L_2$ norm or Euclidean distance measure is used for computing the distance between the images. It is given by the formula,

$$d_{(I1,I2)} = [(f_{I1}-f_{I2})^2]^{1/2} \qquad (1)$$

Where, $f_{I1}$ and $f_{I2}$ are the feature vectors of images $I_1$ and $I_2$.

### 5.1 Minimum distance between images

For computing the minimum distance between the regions of the images a modified Integrated Region Matching algorithm [3] is used. The IRM algorithm allows one region in an image to be matched with several regions of another image. In the proposed algorithm, for each ROI in the query image, the colour and texture features are computed and is compared with each ROIs of the target images (Fig.5). Assume that image $I_1$ has m ROIs represented by $R_1= \{r_1, r_2,\ldots,r_m\}$ and $I_2$ has n ROIs represented by $R_2=\{ r'_1, r'_2,\ldots r'_n\}$. Let the distance between $r_i$ and $r'_j$ be $d(r_i,r'_j)$ denoted as $d_{i,j}$. Every region ri of R1 is compared with every region $r_j$ of $R_2$. This results in 'n' comparisons for a single region in $R_1$ and n distance measures. These distances are stored in ascending order in an array and the minimum distance (d[1]) only is taken for the final computation of the distance D; the distance between $I_1$ and $I_2$. Every d[1] of the 'm' distances is then multiplied with the minimum significance of the corresponding regions. Finally out of the m × n distances m distances are added to get the distance D. Using this method if image $I_1$ is compared with itself, D will be equal to zero indicating perfect match.

The significance matrix $S_1$ and $S_2$ of image $I_1$ and $I_2$ respectively consist of the white pixel density in each identified region. Ie, if $I_1$ has m regions and $I_2$ has n regions,
$S_1 = [s_{11'}, s_{12'}, \ldots s_{1m'}]$ (2)
$S_2 = [s_{21'}, s_{22'}, \ldots s_{2n'}]$ (3)
Where, $s_{1i'}$ and $s_{2i'}$ are the white pixel density in each identified region of $I_1$ and $I_2$. Also, $S_1$ and $S_2$ are normalized so that $\Sigma S_1 = 0$ and $\Sigma S_2 = 0$.

The algorithm is summarized as follows:

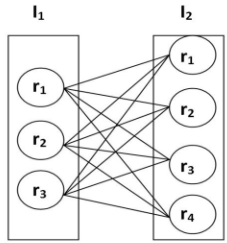

Fig. 5 m regions of $I_1$ are compared with n regions of $I_2$

**Input**: $R_1$, $R_2$; the ROIs of $I_1$ and $I_2$
   $S_1$, $S_2$; significance of selected regions of $I_1$ and $I_2$
**Output**: D, minimum distance between regions of $I_1$ and $I_2$
**Begin**
   **for** each region in the query image $I_1$, i=1 to m **do**
      **for** each region in the target image $I_2$, j=1 to n **do**
         compute distance $d[j] = d_{i,j}$;
      **end**
      Sort distance array 'd' in ascending order;
      **if** ($\Sigma S_1 > 0$ and $\Sigma S_2 > 0$)
         $s'_{i'j'}$ = minimum ($s_{i'}$, $s_{j'}$);
         $D = D + d[1] \times s'_{i'j'}$;
         $s_{i'} = s_{i'} - s'_{i'j'}$;
         $s_{j'} = s_{j'} - s'_{i'j'}$;
      **else**
         $D = D + d[1]$;
      **end if**
   **end for**
**end begin**

'd' is the array containing the distances between the $r_i$ of $R_1$ with the n regions of $R_2$. If d[1] is the minimum distance in the array; the region pair being i of $R_1$ and j of $R_2$, then $s_{i'}$ is the significance of region i in $S_1$, and $s_{j'}$ is the significance of region j in $S_2$ and $s'_{i'j'}$ is the minimum significance among the two.

In some cases $\Sigma S_1$ or $\Sigma S_2$ or both will become zero before all the m regions of the query image $I_1$ is considered for the similarity calculation. In such cases d[1] of the uncounted regions is taken for similarity computation.

The minimum distance between the horizontal and vertical blocks are also computed in a similar manner and is denoted as $D_h$ and $D_v$ respectively. The final distance between $I_1$ and $I_2$ is given by

$D' = D + D_h + D_v + D_g + d_{central\ block\_colour\_texture\_feature}$ (4)

Where, $D_g = d_{global\_colour\_feature} + d_{global\_shape\_feature}$; $d_{global\_colour\_feature}$ and $d_{global\_shape\_feature}$ being the Euclidean distance between the global colour and shape feature vectors of $I_1$ and $I_2$ and $d_{central\ block\_colour\_texture\_feature}$ is the distance between the feature vectors of the central blocks of $I_1$ and $I_2$.

## 6. Experimental Results and Discussions

The Wang's image database [9] of 1000 images, which is considered to be one of the benchmark databases for CBIR, consisting of 10 categories is used for evaluating the performance of the proposed method. Each category contains 100 images. A retrieved image is considered to be correct if and only if it is in the same category as the query. For each query, a preselected number of images are retrieved which are illustrated and listed in the ascending order of the distance between the query and the retrieved images. The results of the proposed method is compared with that of [10], [11] and [27] in terms of average precision. Precision (P) of retrieved results is given by

$P(k) = n_k / k$ (5)

Where, k is the number of retrieved images, $n_k$ is the number of relevant images in the retrieved images. The average precision of the images belonging to the $q^{th}$ category $A_q$ is given by

$$\bar{P}_q = \sum_{k \in A_q} P(I_k) / |A_q|, \quad q = 1, 2, \ldots 10.$$
(6)

The final average precision is

$$\bar{P} = \sum_{q=1}^{10} \bar{P}_q / 10 \qquad (7)$$

Table.1. shows the average precision of the retrieved images for different categories when k=20 for different methods. It is seen that for most of the categories the proposed method provides better or comparable results with that of the other methods. For a few categories like 'Beaches', 'Buildings' and 'Mountains' the performance of the proposed method is lower than that of some of the compared methods because of the similarity of the background of the images. For the categories 'Dinosaur' and 'Flowers' the average precision when k=20 is very high. This means that for images with single object the proposed algorithm works better than the compared algorithms.

Table.1 % Average Precision (K=20) of retrieved images using different methods

| Category | % Average precision of retrieved images for k=20 | | | |
|---|---|---|---|---|
| | *Jhanwar et al[11]* | *Hung and Dai's [10]* | *CTDCIRS [27]* | *Proposed method* |
| Africa | 45.25 | 42.40 | 56.20 | 71.52 |
| Beaches | 39.75 | 44.55 | 53.60 | 43.60 |
| Buildings | 37.35 | 41.05 | 61.00 | 53.55 |
| Bus | 74.10 | 85.15 | 89.30 | 85.30 |
| Dinosaur | 91.45 | 58.65 | 98.40 | 99.55 |
| Elephant | 30.40 | 42.55 | 57.80 | 59.10 |
| Flowers | 85.15 | 89.75 | 89.90 | 90.95 |
| Horse | 56.80 | 58.90 | 78.00 | 92.40 |
| Mountains | 29.25 | 28.5 | 51.20 | 38.35 |
| Food | 36.95 | 42.65 | 69.40 | 72.40 |
| **Average** | **52.64** | **53.24** | **70.48** | **70.67** |

Fig.6 depicts the top 19 retrieved images for two sample query image using proposed method. In each set, on top left corner is the query image and the retrieved images are listed according to their distance with the query image.

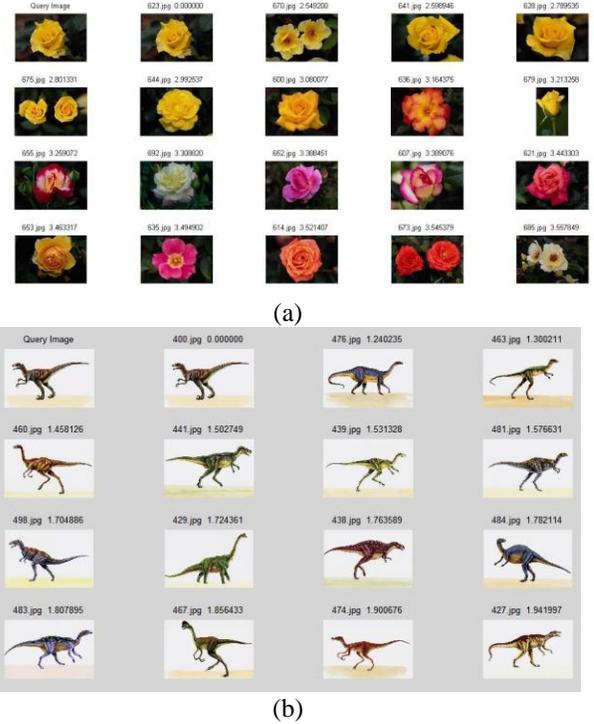

(a)

(b)

Fig. 6 Top 19 retrieved images for the two sample query image. For both the results the image in the top left corner is the query image and the retrieved images are listed according to their distance from the query image.

## 6. Conclusion and future work

A content based image retrieval system using the colour and texture features of selected sub-blocks and global colour and shape features of the image is proposed. The colour features are extracted from the histograms of the quantized HSV color space, texture features from GLCM and shape features from EHD. A modified IRM algorithm is used for computing the minimum distance between the selected sub-blocks of the query image and the candidate images in the database. Unlike the most sub-block based methods that involves all the sub-blocks of the query image to be compared with that of the candidate images, our system involves only selected sub-blocks for similarity measurement, thus reducing the number of comparisons and computational cost. Experimental results also show that the proposed method provides better retrieving result than some of the existing methods. Future work aims at the selection of sub-blocks based on their saliency in the image to improve the retrieval precision. Also the proposed method has to be tested on various databases to test the robustness.

**E R Vimina**, Department of Computer Science, Rajagiri College of Social Sciences, India received B.Tech in Electrical and Electronics Engineering from Mahatma Gandhi University, India and ME degree in Computer Science and Engineering from Bharathiar University, India. She is currently pursuing doctoral research in Image Retrieval at Cochin University of Science and Technology, India. Her research interests include Image Retrieval and Artificial intelligence.

**Dr. K. Paulose Jacob**, Professor of Computer Science at Cochin University of Science and Technology since 1994, is the Director of the School of Computer Science Studies and Dean of the Faculty of Engineering. Dr. Jacob has been teaching at the Cochin University since 1980. A National Merit Scholar all through, he is a graduate in Electrical Engineering and postgraduate in Digital Electronics. He obtained Ph D in Computer Engineering for his work in Multi-Microprocessor Applications. His other research interests are Information Systems Engineering, Intelligent

Architectures and Networks. He has more than 50 research publications to his credit, and has presented research papers in several International Conferences in Europe, USA, UK and other countries. He is a Permanent Professional Member of the ACM and a Life Member of the Computer Society of India.